\documentclass{article}
\usepackage{epsfig}

\usepackage{a4}
\newcommand{\beq}{\begin{equation}}
\newcommand{\eeq}{\end{equation}}
\newcommand{\beqs}{\begin{eqnarray}}
\newcommand{\eeqs}{\end{eqnarray}}
\newcommand{\ra}{\rightarrow}

\usepackage{amsmath}

\begin{document}
\title{Features of gravity-Yang-Mills hierarchies \\ in $d-$dimensions}
 \author{{\large Eugen Radu}$^{\dagger}$,
{\large Cristian Stelea}$^{\ddagger}$ 
and {\large D. H. Tchrakian}$^{\dagger \star}$ \\ \\
$^{\dagger}${\small Department of
Mathematical Physics,  National University of Ireland, Maynooth, Ireland} \\
$^{\ddagger}${\small Department of Physics, University of Waterloo,
Ontario N2L 3G1, Canada}\\  
$^{\star}${\small School of Theoretical Physics -- DIAS, 10 Burlington
Road, Dublin 4, Ireland }}
\date{}
\newcommand{\dd}{\mbox{d}}
\newcommand{\tr}{\mbox{tr}}
\newcommand{\la}{\lambda}
\newcommand{\ka}{\kappa}
\newcommand{\al}{\alpha}
\newcommand{\ga}{\gamma}
\newcommand{\de}{\delta}
\newcommand{\si}{\sigma}
\newcommand{\bomega}{\mbox{\boldmath $\omega$}}
\newcommand{\bsi}{\mbox{\boldmath $\sigma$}}
\newcommand{\bchi}{\mbox{\boldmath $\chi$}}
\newcommand{\bal}{\mbox{\boldmath $\alpha$}}
\newcommand{\bpsi}{\mbox{\boldmath $\psi$}}
\newcommand{\brho}{\mbox{\boldmath $\varrho$}}
\newcommand{\beps}{\mbox{\boldmath $\varepsilon$}}
\newcommand{\bxi}{\mbox{\boldmath $\xi$}}
\newcommand{\bbeta}{\mbox{\boldmath $\beta$}}
\newcommand{\ee}{\end{equation}}
\newcommand{\eea}{\end{eqnarray}}
\newcommand{\be}{\begin{equation}}
\newcommand{\bea}{\begin{eqnarray}}
\newcommand{\ii}{\mbox{i}}
\newcommand{\e}{\mbox{e}}
\newcommand{\pa}{\partial}
\newcommand{\Om}{\Omega}
\newcommand{\vep}{\varepsilon}
\newcommand{\bfph}{{\bf \phi}}
\newcommand{\lm}{\lambda}
\def\theequation{\arabic{equation}}
\renewcommand{\thefootnote}{\fnsymbol{footnote}}
\newcommand{\re}[1]{(\ref{#1})}
\newcommand{\R}{{\rm I \hspace{-0.52ex} R}}
\newcommand{\N}{{\sf N\hspace*{-1.0ex}\rule{0.15ex}%
{1.3ex}\hspace*{1.0ex}}}
\newcommand{\Q}{{\sf Q\hspace*{-1.1ex}\rule{0.15ex}%
{1.5ex}\hspace*{1.1ex}}}
\newcommand{\C}{{\sf C\hspace*{-0.9ex}\rule{0.15ex}%
{1.3ex}\hspace*{0.9ex}}}
\newcommand{\eins}{1\hspace{-0.56ex}{\rm I}}
\renewcommand{\thefootnote}{\arabic{footnote}}

\maketitle


\medskip

\begin{abstract}
Higher dimensional, direct analogues of the usual $d=4$ Einstein--Yang-Mills
(EYM) systems are studied. These consist of the gravitational and Yang-Mills
hierarchies in $d=4p$ dimensional spacetimes, both consisting of $2p$-form
curvature terms only. Regular and black hole solutions are constructed in
$2p+2\le d \le 4p$, in
which dimensions the total mass-energy is finite, generalising the familiar
Bartnik-McKinnon solutions in EYM theory for $p=1$. In $d=4p$, this similarity
is complete. In the special case of $d=2p+1$, just beyond the finite energy
range of $d$, exact solutions in closed form are found. Finally, $d=2p+1$
purely gravitational systems, 
whose solutions generalise the static $d=3$ BTZ solutions,
are discussed.

\end{abstract}

\medskip
\section{Introduction}
Regular and black hole solutions of gravitating gauge field systems have
been studied for a long time since the pioneering work of Bartnik and McKinnon
(BK) \cite{BK}, where regular solutions of the Yang--Mills (YM) and
Einstein--Hilbert (EH) systems were presented for  $d=4$ spacetime dimensions.
Other models of gravitating non-Abelian gauge fields usually
share a number of common features with the BK particles, 
this example becoming canonical.
After that discovery, there has been a great deal of numerical and analytical 
work on various aspects of Einstein-Yang-Mills (EYM) theory and a variety of
self-gravitating structures with non-Abelian fields have been found (for a
review see \cite{Volkov:2001tb}).
These include hairy black holes solutions, which led to
certain revisions of some of the basic concepts of black hole physics
based on the uniqueness and no-hair theorems.

In the last years, gravitating solutions with nonabelian fields enjoyed
renewed interest
in the AdS/CFT correspondence, and potentially in the study of $Dp$-branes
of superstring theory. It is therefore pertinent to extend the study of
such solutions to higher dimensions other than that of $d=4$ spacetime.

Recently regular solutions were found in $d=6,7,8$ in \cite{BCT}; both regular
and black hole solutions in five dimensions were presented in \cite{BCHT},
the $d=6,7,8$ black hole solutions being discussed in \cite{Radu:2005mj}.
These were the classical solutions to systems consisting of higher order
terms in both the
YM and the gravitational curvatures, which do in fact appear
in the low energy effective action of string theory. Such terms employed in
\cite{BCT,BCHT} were those constructed from the totally antisymmetrised
$2p$-forms in both the YM and the Riemann curvature $2$-forms, namely the YM
and the gravitational hierarchies labelled by integers $p$. The $p=1$ members
in each case are the usual YM and the EH systems respectively, while the $p=2$
gravitational member is the familiar Gauss-Bonnet term.

These asymptotically flat configurations differ in many respects from the
$d=4$ BK solutions. In particular, all $d>4$  solutions have 
only one node in the gauge potential. Also, the black holes exist only up 
to a maximal value of the event horizon radius.

The salient property of these solutions~\cite{BCT,BCHT} is that they exist only
for a limited range of the coupling parameter $\al^2$ (which gives the
strength of  the gravitational interaction), and exhibit critical
behaviours in $\al^2$. This is not surprising since such composite models
necessarily feature more than one dimensional constant, analogous to
gravitating monopoles \cite{BFM1}. (In the latter case dimensionful Higgs
vacuum expectation value (VEV) plays this role.)

More recently, a complete analysis of these critical behaviours in $\al^2$ in
the higher dimensional gravitating YM systems was presented in \cite{BMT}. The
methods used in \cite{BMT} were those of fixed point analysis developed
previously in \cite{BFM2} for proving the existence of the BK 
solutions ~\cite{BK} to the usual EYM system in $d=4$.
Concerning the fairly complicated landscape of critical points here~\cite{BMT},
we restrict our comments only to pointing out that in addition to the two types
of critical points occurring in gravitating monopoles~\cite{BFM1}, namely the
ones associated with the {\it end point} and the {\it Reissner-Nordstr{\o}m}
types, there is also a {\it conical} fixed point, so described since at that
scaling point an angular deficit appears in the gravitational metric function.
We refer the reader to \cite{BMT} for a full
account of these critical behaviours.

The analysis in \cite{BMT} was carried out for models whose gravitational part
consisted only of the $p=1$ member of the gravitational hierarchy, namely the
Einstein--Hilbert (EH) term. The physical reason for this was that the $p\ge 2$
Gauss-Bonnet like terms play only a quantitative role, confirmed numerically
in \cite{BCT}, and have no effect on the existence of finite mass solutions.
By contrast in the gauge field sector, various combinations of YM
terms with $p\ge 1$ essential for the existence of such solutions were
employed, as required by scaling arguments. Hence higher $p\ge 1$ YM terms
were employed there, once $d$ was greater than $4$. Thus the analysis in
\cite{BMT} probed the effect of the higher $p\ge 2$ YM terms.

One of the two main aims of the present work is to probe the effect of higher
order $p\ge 2$ gravitational terms in EYM models. But we know from the
numerical results of \cite{BCT} that once the $p=1$ (EH) term is present, the
effects of all higher $p\ge 2$ terms become masked. It is therefore the case
that if one wishes to study the effects of the $p$-th gravitational term, all
other gravitational terms with $p_i<p$ must be excluded. In practical terms,
this means that we will restrict to models featuring only the $p$-th
gravitational term.

The other one of our two main aims is to exhibit certain qualitative
similarities of the solutions supported by a family of EYM models, the first
member of which consists of the $p=1$ gravitational (EH) and the $p=1$ YM
terms, supporting the BK solutions. It follows naturally that the YM term we
must choose interacting with the (unique) $p$-th gravitational term, is the
(unique) $p$-th YM term, in spacetime dimensions $d=4p$. We expect to exhibit
a recurring `symmetry' in the properties of the solutions of this family of
models {\it modulo} $4p$ dimensions. In the models studied here
the complicated critical features of the
gravitating monopole \cite{BFM1} and of the higher dimensional EYM solutions
studied in \cite{BCT,BCHT,BMT} will be absent, since these are due to the
presence of at least one additional (to the
gravitational and YM couplings) dimensionful constant, e.g. the Higgs
VEV, or, the higher curvature YM coupling constants.

In Section 2 we introduce the relevant gravitating YM models, and subject
them to spherical symmetry. These models will be characterised by two equal
integers $p_1=p_2=p$ specifying the model and the gauge group, and the
dimension of the spacetime $d$. In Section 3 we specialise the dimension $d$
to the values $d=4p,\,d=4p-1,\,...,d=2p+2$, which are the only dimensions
in which asymptotically flat finite energy solutions exist. These solutions
can be constructed only numerically so the results presented in Section 3 are
mainly numerical. In Section 4 we specialise to spacetime dimensions $d=2p+1$
in which no asymptotically flat, finite energy solutions exist for $p_1=p_2=p$.
The interest in the latter, inspite
of their pathological properties, is that they can be given in
closed form, which is a novel feature in gravitating gauge field theory.
In addition to these we have supplied two Appendices, A and B, devoted to the
extension of some the models studied in Sections 3 and 4, to feature a
non-vanishing cosmological constant $\Lambda$. Like the solutions presented in
Section 3, those in Appendices A and B are expressed in closed form.
In Appendix A we study the geometric properties of the gauge decoupled limit 
models in $d=2p+1$ appearing in Section 3, supplemented with a
cosmological term, which
are the hierarchy of gravity solutions pertaining to the 
static BTZ solution~\cite{Banados:1992wn}, this
last being the $p=1$ member. The Appendix B presents a generalisation of the
exact gravity-YM $p_1=p_2=p$ solution in $d=2p+1$ dimensions for a nonzero
$\Lambda$. In Section 5, we summarise our results.

\section{The model and imposition of spherical symmetry}
The precise form of the gravitational and non-Abelian matter content of string
effective actions beyond leading order is still an evolving research
subject~\cite{PVW}. Our
position in studying higher dimensional gravitating YM solutions, in
\cite{BCT,BCHT,BMT}, has been to choose what we have referred to as the
superposition of the $p$-th members of the
gravitational and YM hierarchies, consisting of all possible higher order
curvature forms allowed in any given dimension.

The said gravitational system is the superposition of all
possible $(p,q)$-Ricci scalars $R_{(p,q)}$ 
\be
\label{EHhier}
{\cal L}_{\rm{grav}}^{(P)}\ =\ \sum_{p=1}^{P}\
\frac{\kappa_{p}}{2p}\ e \ R_{(p,q)}\ ,
\ee
where $R_{(p,q)}$ are constructed from the $2p$-form $R(2p)=R\wedge R\wedge
...\wedge R$ resulting from the totally antisymmetrised $p$-fold
products of the Riemann curvature $2$-forms $R$. We express
$R_{(p,q)}$ in the notation of \cite{O'Brien:1988rs} as
\be
\label{gp}
e\,R_{(p,q)}=\vep^{\mu_1\mu_2...\mu_{2p}\nu_1\nu_2...\nu_{q}}
e_{\nu_1}^{n_1}e_{\nu_2}^{n_2}...e_{\nu_q}^{n_q}
~\vep_{m_1m_2...m_{2p}n_1n_2...n_q}\,
R_{\mu_1\mu_2....\mu_{2p}}^{m_1m_2..m_{2p}}\,,
\ee
where $e_{\nu}^{n}$ are the {\it Vielbein} fields,
$e=\mbox{det}(e_{\nu}^{n})$ in \re{EHhier}, and
$R_{\mu_1\mu_2....\mu_{2p}}^{m_1m_2..m_{2p}}=R(2p)$ is the $p$-fold totally
antisymmetrised product of the Riemann curvature, in component notation. 
This leads to the definition of the $p$-th Einstein tensor
\be
\label{peinstein}
G_{(p)}{}_{\mu}^a=R_{(p)}{}_{\mu}^a\
-\ \frac{1}{2p}\ e_{\mu}^a\ R_{(p)}\ .
\ee
One reads from \re{gp}
that
\[
d=2p+q\,.
\]
Now the minimum nontrivial value of $q$ is $q=1$, since when $q=0$ \re{gp}
is manifestly a total divergence, namely the Euler-Hirzebruch density. Thus
the highest nontrivial value of $P$ in the superposition \re{EHhier} is
$$P_{\rm{max}}\le\frac12(d-1).$$

The corresponding superposition of the members of the YM hierarchy is
\be
\label{YMhier}
{\cal L}_{\rm{YM}}^{(P)}=\sum_{p=1}^{P}\frac{\tau_p}{2(2p)!}\ e\
\mbox{Tr}\,F(2p)^2\;,
\ee
in which the $2p$-form $F(2p)$ is the $p$-fold antisymmetrised product
$F(2p)=F\wedge F\wedge...\wedge F$ of the YM curvature $2$-form $F$. Here
the maximum value of $P$ in the superposition \re{YMhier} is simply
$P_{\rm{max}}\le\frac12d.$
We define the $p$-stress tensor pertaining to each term in (\ref{YMhier}) as
\be
T_{\mu\nu}^{(p)}=
\mbox{Tr}\ F(2p)_{\mu\lambda_1\lambda_2...\lambda_{2p-1}}
F(2p)_{\nu}{}^{\lambda_1\lambda_2...\lambda_{2p-1}}
-\frac{1}{4p}g_{\mu\nu}\ \mbox{Tr}\ F(2p)_{\lambda_1\lambda_2...\lambda_{2p}}
F(2p)^{\lambda_1\lambda_2...\lambda_{2p}}\ .
\label{pstress}
\ee
The $P=1$ systems \re{EHhier} and \re{YMhier} are the usual EH gravity and
YM theories, respectively.

In \cite{BCT,BCHT,BMT}, some convenient superpositions \re{EHhier} and
\re{YMhier} were selected to be studied, taking into account the particular
properties of the solutions that were being sought.

It is our aim here to truncate both \re{EHhier} and \re{YMhier} such that
only one term is present in each. This is so that, analogously to the
 BK  case, the solutions not be parameterized by one (e.g.
$\al^2$) or more parameters. The other criterion stated in Section 1 is that in
the appropriate dimensions higher members of the gravitational hierarchy be
employed. So far, subject to respecting the Derrick scaling requirements, one
can choose any $P=p_1$ in \re{EHhier} and any $P=p_2$ in \re{YMhier}. The
final criterion is that of symmetry and analogy with the BK case in $d=4$,
namely that $p_1=p_2=p$ in $d=4p$
dimensions~\footnote{There is yet another apparently
unrelated coincidence here. In
Euclidean signature, the double-self-duality of the Riemann $2p$ form curvature
leads to the vacuum $p$-Einstein equations being satisfied, and yields
a self-dual $SO_{\pm}(4p)$ YM $2p$-form field strength.}.

Thus we define the gravitating YM models in $2p+2\le d\le 4p$ spacetime
dimensions, whose static finite energy solutions will be
constructed numerically in the next section, by the Lagrangians
\be
\label{lagp}
{\cal L}_{(p,d)}=e\,\left(\frac{\kappa_{p}}{2p}\ \ R_{(p,q)}
+\frac{\tau_p}{2(2p)!}\ \mbox{Tr}\,F(2p)^2\right)\quad,\quad
2\le q\le 2p\;,
\ee
$\kappa_{p}$ and $\tau_p$ being constants giving 
the strength of the gravitational and YM interactions, respectively.
Note that the gravitational part in \re{lagp} is described by the
$(p,q)$-Ricci scalars $R_{(p,q)}$ in which $q=d-2p$, as defined in
\re{gp}.
For this system, the variational equations for the YM and gravitational
fields are 
\be
\label{vareq}
\tau_pD_{\mu}\left(e\ F^{\mu\nu}\right)=0,
~~~
\ka_pG_{(p)}{}_{\mu}^a\ =\
\frac{\tau_p}{2(2p)!} T^{(p)}{}_{\mu }^a .
\ee

The model \re{lagp} directly generalises the usual EYM model in $d=4$
spacetime for $d=4p$, the latter being the $p=1$ case. In that case the
dimensions of the gravitational part of \re{lagp} are $L^{-2p}$, versus the
dimensions of the YM part $L^{-4p}$. This choice has been made on grounds of
symmetry rather than physics, since the leading terms in the effective action
of string theory are the $p=1$ EH and YM terms, which are both excluded.
Since the $p=1$ EYM system is best
known for its  BK  solutions, the solutions to the
hierarchy defined by \re{lagp} for $d=4p$ might be described as the BK
hierarchy.

The definition of the model becomes complete on specifying the gauge groups
and their representations. Adopting the criterion of employing chiral
representations, both for {\it even} and for {\it odd} $d$ in \re{lagp}, it is
convenient to choose the gauge group to be $SO(\bar d)$. We shall therefore
denote our representation
matrices by $SO_{\pm}(\bar d)$, where $\bar d=d$ and $\bar d=d-1$ for
{\it even} and {\it odd} $d$ respectively.

In this unified notation (for odd and even $d$), the spherically symmetric
Ansatz for the $SO_{\pm}(\bar d)$-valued gauge fields then reads
\cite{BCT,BCHT}
\begin{equation}
\label{YMsph}
A_0=0\ ,\quad
A_i=\left(\frac{1-w(r)}{r}\right)\Sigma_{ij}^{(\pm)}\hat x_j\ , \quad
\Sigma_{ij}^{(\pm)}=
-\frac{1}{4}\left(\frac{1\pm\Gamma_{ \bar d+1}}{2}\right)
[\Gamma_i ,\Gamma_j]\ .
\end{equation}
The $\Gamma$'s denote the $\bar d$-dimensional gamma matrices and
$1,~j=1,2,...,d-1$ for both cases.

The spherically symmetric
metric Ansatz we use is parameterized by two functions $N(r)$ and $\sigma(r)$
\be
\label{metric}
ds^2 = - N(r)\sigma^2(r) dt^2\ +\ N(r)^{-1} dr^2\ 
+\ r^2 d \Omega_{(d-2)}^2 \: .
\ee

Inserting \re{YMsph} in \re{lagp}, the resulting reduced one dimensional
Lagrangian  is
\bea
L_{(p,d)}&=&
\frac{(d-2)!}{(d-2p-1)!}\si\,\Bigg\{\frac{\ka_p}{2^{2p-1}}
\frac{d}{dr}\left[r^{d-2p-1}(1-N)^p\right]\nonumber\\
&+&r^{d-2}\frac{\tau_p}{2\cdot (2p)!}\,W^{p-1}
\left[(2p)N\left(\frac1r\frac{dw}{dr}\right)^2+
(d-2p-1)\,W\right]\Bigg\},
\label{redlag}
\eea
where we note
\be
\label{W}
W=\frac{(1-w^2)^2}{r^4}.
\ee

\section{A Bartnik-McKinnon hierarchy}

 It follows from the  Derrick-type scaling
arguments that static finite energy solutions to the field equations of the
model \re{lagp} exist only in spacetime dimensions
\be
\label{virial}
2p+2\le d\le 4p\,,
\ee
and can only be constructed numerically, which we present here. The $d=4p$
family is referred to as the BK hierarchy, but we study all possible cases
allowed by \re{virial}. Note that for $p=1$, $d=4$ 
is the only possibility allowed by \re{virial}.

\newpage
\setlength{\unitlength}{1cm}

\begin{picture}(18,7)
\label{fig1}
\centering
\put(2,0.0){\epsfig{file=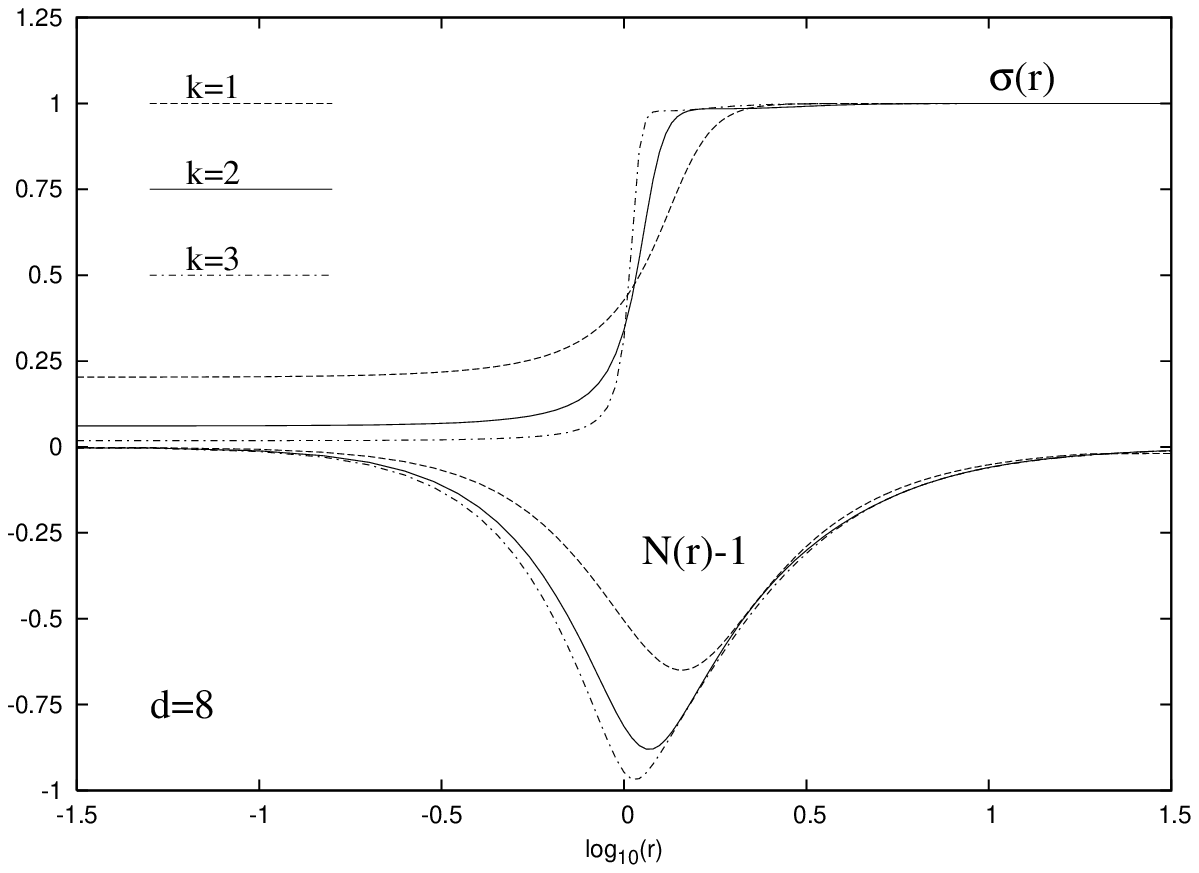,width=11cm}}
\end{picture}
\begin{picture}(19,8.)
\centering
\put(2.6,0.0){\epsfig{file=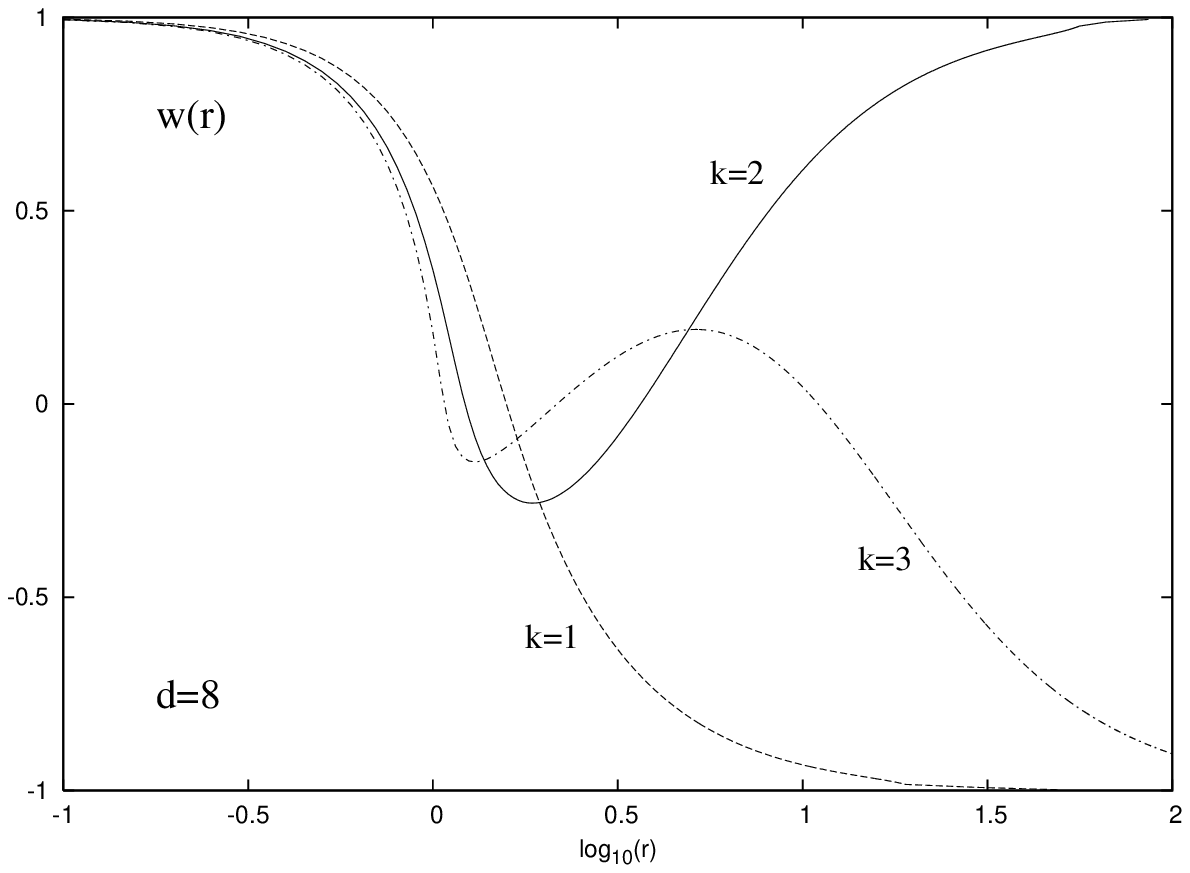,width=11cm}}
\end{picture}
\\
\\
{\small {\bf Figure 1.}  
The profiles of the metric functions $N(r)$, $\si(r)$
and gauge function $w(r)$
 are presented for $k-$node globally regular solutions of the $p=2$ 
gravity-Yang-Mills model in $d=8$ dimensions. }
\\
\\
When looking for numerical solutions, it is 
convenient to redefine 
\[
r \to  \left(\frac{2^{2(p-1)}\tau_p}{(2p!)\kappa_p}\right)^{1/(4p-2)} r
\]
such that  for any $(d,~p)$ no free parameter appears in the field equations.

The field equations implies the relations
\begin{eqnarray}
\label{eq1}
\frac{d}{dr}\big[r^{d-2p-1}(1-N)^p\big]&=&r^{d-2}W^{p-1}
\bigg[2pN \left(\frac{1}{r}\frac{dw}{dr}\right)^2+(d-2p-1)W \bigg],
\\
\label{eq2}
\frac{d\sigma }{dr}&=&2r^{2p-3}\sigma(1-N)^{1-p}W^{p-1}w'^2,
\end{eqnarray}

\newpage
\setlength{\unitlength}{1cm}

\begin{picture}(18,7)
\label{fig2}
\centering
\put(2,0.0){\epsfig{file=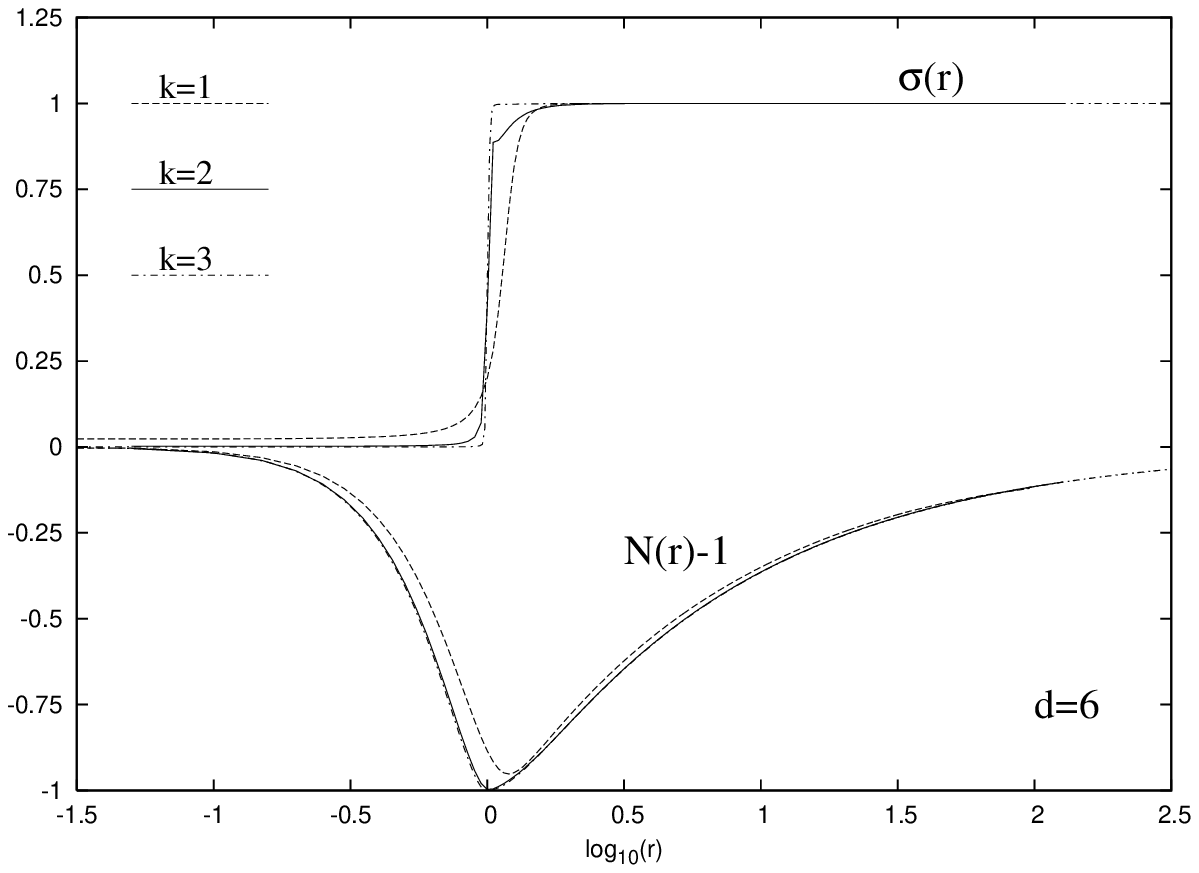,width=11cm}}
\end{picture}
\begin{picture}(19,8.)
\centering
\put(2.6,0.0){\epsfig{file=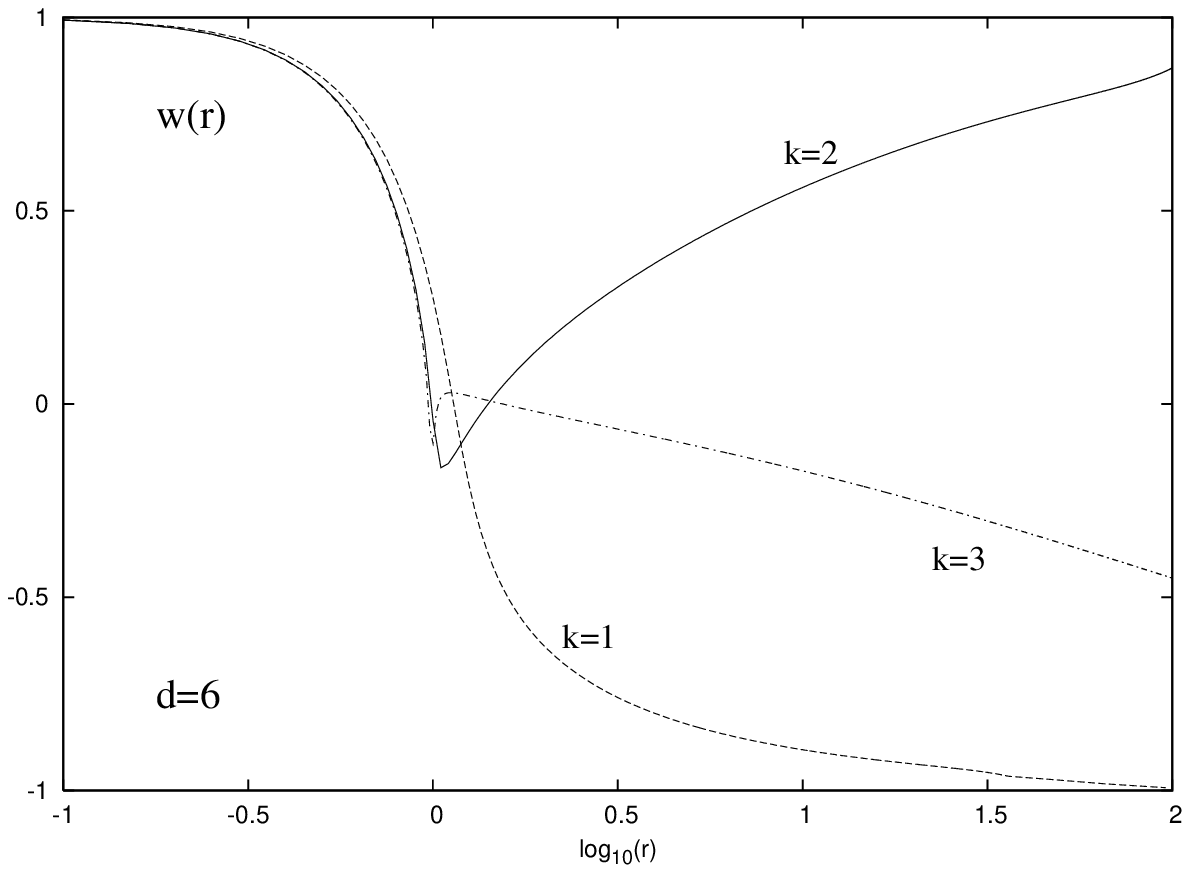,width=11cm}}
\end{picture}
\\
\\
{\small {\bf Figure 2.} 
One, two and three nodes solutions of the $p=2$ 
gravity-Yang-Mills model in $d=6$ dimensions. }
\\
\\
for the metric function, and
\begin{eqnarray}
\label{eq3}
\frac{d}{dr}\left(r^{d-4}\sigma W^{p-1}N w'\right)&=&
\sigma r^{d-6}W^{p-2}w(w^2-1)\left(2(p-1)N(\frac{1}{r}\frac{dw}{dr})^2
+(d-2p-1)W\right).
\end{eqnarray}
for the gauge potential.
 
For globally regular solutions, finite energy requirements and regularity of 
the metric at $r=0$ give
\begin{eqnarray}
\label{as2} 
w=1-br^2+O(r^4),~~N=1-4b^2 r^2+O(r^3),~~\sigma=\sigma_0(1+4b^2 r^2)+O(r^4),
\end{eqnarray}
where $\sigma_0,~b$ are two positive constant.
The analysis of the field equations as $r\to\infty$ gives
\begin{eqnarray}
\label{funct-asympt}
N=1- \frac{M_0^{1/p}}{r^{(d-2p-1)/p}}+\dots,
~~ 
w=\pm 1+\frac{c}{r^{(2p-d+1)/p}}+\dots,
\\
\nonumber
\sigma=1+\frac{2p(4c^2)^pM_0^{(1-p)/p}}{(d-2p-1)(3p-1)
+2p^2}\frac{1}{r^{2p+(d-2p-1)(3p-1)/p}}+\dots,
\end{eqnarray}

\newpage
\setlength{\unitlength}{1cm}

\begin{picture}(18,7)
\label{fig3}
\centering
\put(2,0.0){\epsfig{file=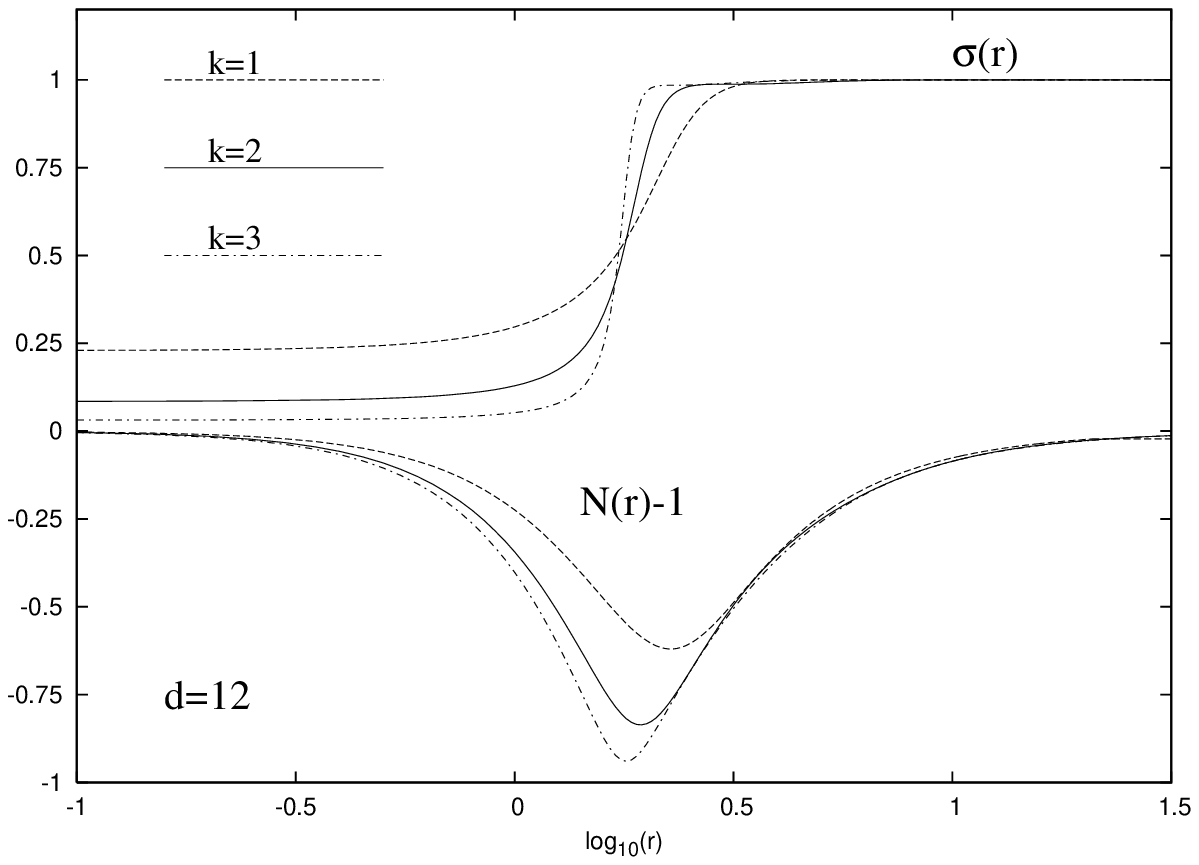,width=11cm}}
\end{picture}
\begin{picture}(19,8.)
\centering
\put(2.6,0.0){\epsfig{file=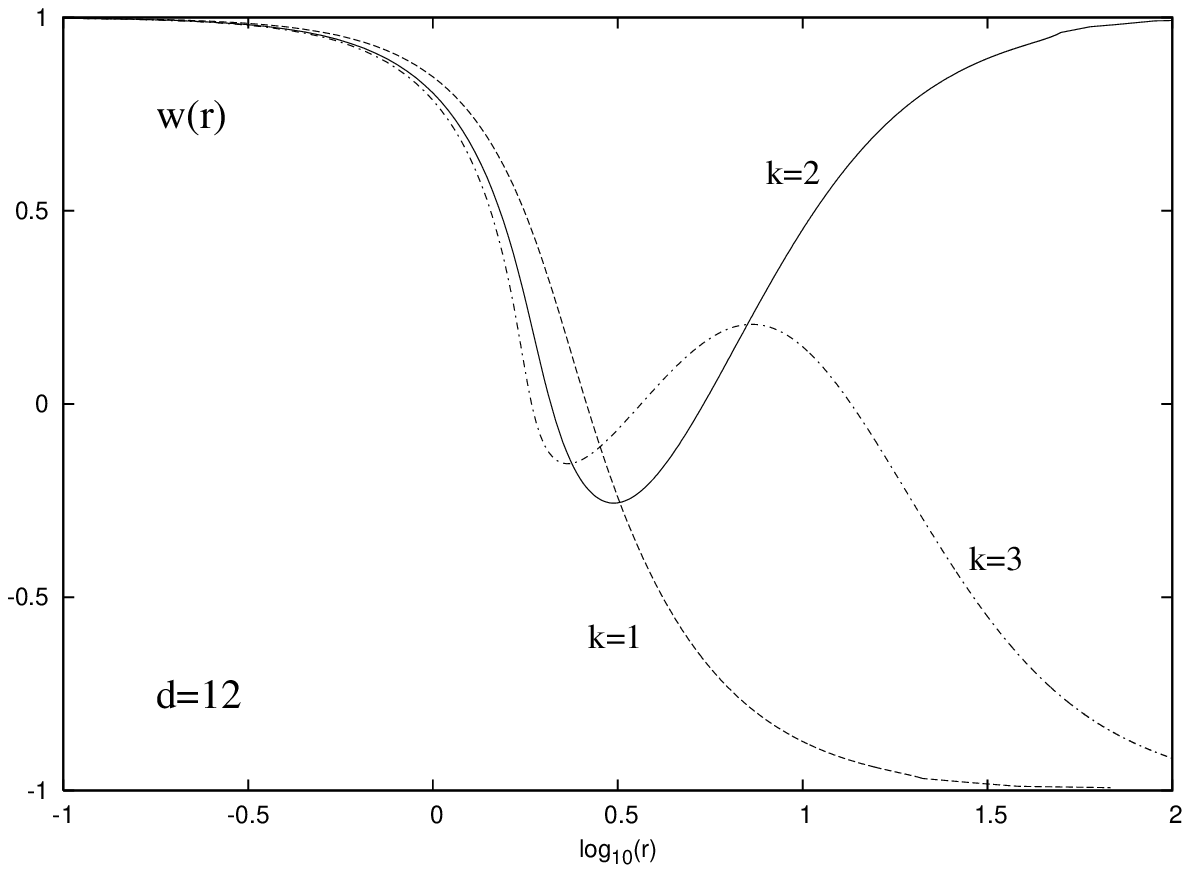,width=11cm}}
\end{picture}
\\
\\
{\small {\bf Figure 3.} The profiles for the metric functions $N(r)$, $\si(r)$
and gauge function $w(r)$
are presented for $k-$node globally regular solutions for the $p=3$ 
gravity-Yang-Mills system in $d=12$ dimensions. }
\\
\\
with $M_0$, $c$ arbitrary constants.
In analogy to the Einstein gravity, it is natural to identify the constant
$M_0$ that enters the asymptotic of $N$ with the mass $\bf{M}$ of the
solutions, up to a $d-$dependent factor. To see this let us notice that
(\ref{eq1}) can be written as
\beqs
\frac{d}{dr}[r^{d-2p-1}(1-N)^p]&=&r^{d-2}T^{(p)t}_{~~~t}
\eeqs
which in the $(p=1,d=4)$ case of standard Einstein gravity leads to the
usual definition of mass. Here $T^{(p)t}_{~~~t}$ is the
`energy density' component of the stress-energy tensor defined in
(\ref{pstress}). By keeping with this analogy, `the total energy' inside a
$(d-2)$-sphere $S_{d-2}$ with radius $r$ will be given by
\beqs
\int_0^r dr' \int_{S_{d-2}}d\Omega_{(d-2)}~ r'^{d-2}W^{p-1}(r')
\bigg[2pN(r') \left(\frac{1}{r'}\frac{dw}{dr'}\right)^2+(d-2p-1)W(r') \bigg]
=Area(S_{d-2})r^{d-2p-1}(1-N)^p
\eeqs
\newpage
\setlength{\unitlength}{1cm}

\begin{picture}(18,7)
\label{fig4}
\centering
\put(2,0.0){\epsfig{file=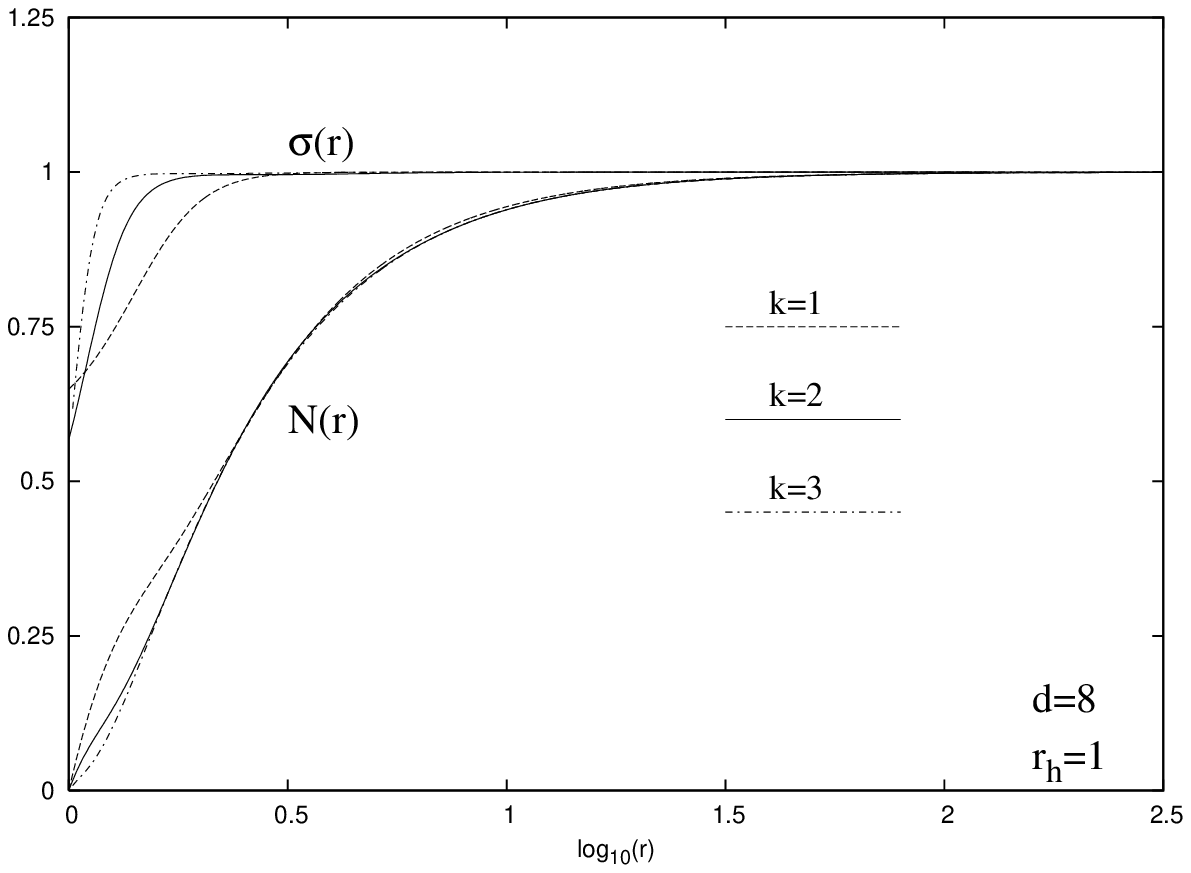,width=11cm}}
\end{picture}
\begin{picture}(19,8.)
\centering
\put(2.6,0.0){\epsfig{file=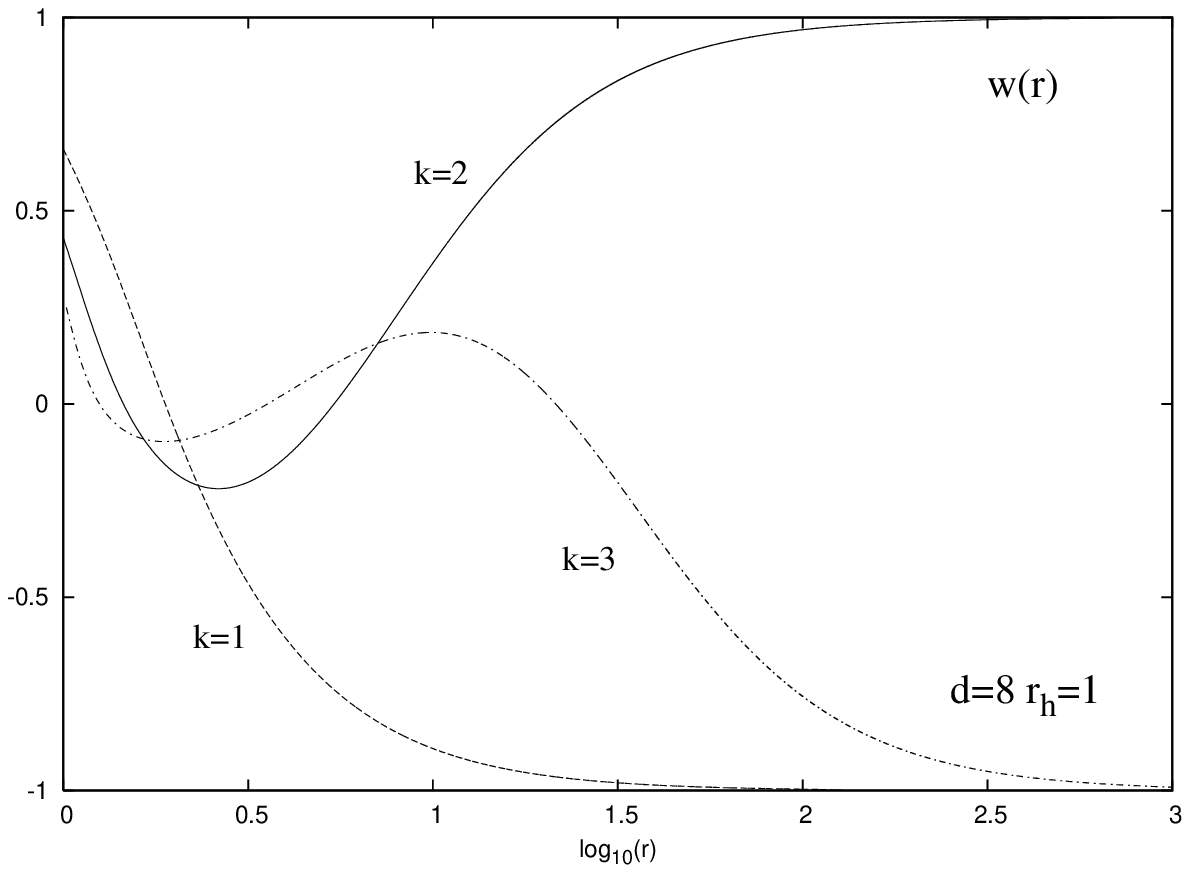,width=11cm}}
\end{picture}
\\
\\
{\small {\bf Figure 4.}
 One, two and three nodes black holes solutions of the $p=2$ 
gravity-Yang-Mills model in $d=8$ dimensions. }
\\
\\
where we made use of the regularity of the function $N(r)$ at origin $r=0$
(see (\ref{as2})). Taking now the limit $r\ra\infty$ we find that the total
energy is finite and given precisely by $\bf{M}$=$Area(S_{d-2})M_0$ as claimed.

The equations of motion (\ref{eq1})-(\ref{eq3})  have been solved
for $p=2,~3$ and a range of $d$.
Considering first globally regular
configurations, we follow the usual approach and, by using a standard ordinary
differential equation solver, we evaluate  the  initial  conditions (\ref{as2})
at $r=10^{-4}$ for global  tolerance $10^{-12}$, adjusting  for fixed shooting
parameter $b$ and  integrating  towards  $r\to\infty$.

As expected, these solutions have many features in
common with the well-known $d=4,~p=1$ BK solutions.
By adjusting the free parameter $b$ appearing in the expansion
(\ref{as2}), we "shoot" for global solutions with the right asymptotics.
The solutions are indexed by $k$-the number of nodes of the gauge potential
$w$, which is always bounded within the strip $|w(r)|\leq 1$.

For the $k$-th solution, the function $w(r)$ has $k$ nodes in the
interval $0<r<\infty$, such that $w(\infty)=(-1)^k$. Typical profiles for
$p=2$ solutions are plotted in Figures 1,~2 (for $d=8,~6$) 
and $p=3,~d=12$ configurations  (Figure 3). It is worth noting here that for a
$p$-gravity
and $p$-YM theory, when the spacetime dimension is $4p$, then the profiles of
the functions $N(r)$, $\si(r)$ and $w(r)$ are perfectly smooth as seen from
Figures 1 (for $d=8$) and Figures 3 (for $d=12$), as is the case for the BK
solutions in $d=4$, while for dimensions different from $d=4p$, e.g. for $d=6$
depicted in Figure 2, these profiles are less smooth and the numerical analysis
is correspondingly more delicate. This is one the features highlighting the
similarity of the solutions to these theories {\it modulo} $4p$.

The behaviour of the metric functions $N$ and $\sigma$ is similar for all
$k$'s. The metric functions $\sigma$ increases with growing 
$r$ from $\sigma(0)=\sigma_0>0$ at the origin to $\sigma(\infty)=1$.
As $k$ increases, $N$ develops a more and more deep minimum at some $r_m$,
closely approaching  the zero value for large enough values of $k$, as
indicated in Figures 1-3. At the same time, the value at the origin of the
metric function $\sigma$ strongly decreases with $k$.

The limiting solution with $k=\infty$ can also be investigated \cite{maison}.
Since the Schwarzschild coordinate system breaks down in this limit, 
we should use a different parameterization and the equations are formulated as
a dynamical system.
Similar to the $p=1$ case, this turns out to be non-asymptotically flat.

The system (\ref{eq1})-(\ref{eq3}) presents also black hole solutions.
The boundary conditions at infinity are still given by (\ref{funct-asympt}), 
and are now supplemented with the requirement that there is a regular 
event horizon at some 
$r=r_h>0$, with $N(r)>0$ for $r>r_h$.
The local power series in the vicinity of the horizon reads
\begin{eqnarray}
\label{as3} 
w(r)=w_h+w'_h(r-r_h)+O(r-r_h)^2,~N(r)=N'_h(r-r_h)+O(r-r_h)^2,~
\sigma(r)=\sigma_h+\sigma'_h(r-r_h)+O(r-r_h)^2,
\end{eqnarray}
with
\begin{eqnarray}
\label{as4} 
N'_h=\frac{(d-2p-1)}{r_h}(1-r^{2p}W^p_h),~~
w'_h= w_h(w_h^2-1)(d-2p-1)\frac{1}{r^2N'_h},~~
\sigma'_h=2\sigma_hr_h^{2p-3}W_h^{p-1}w'^{2}_h,
\end{eqnarray}
and $W_h=\frac{(1-w_h^2)^2}{r_h^4}$.
Similar to the $d=4,~p=1$ case,  one find a sequence 
of global solutions in the interval
$r_h<r<\infty$, for any value of the event horizon $r_h>0$.
These solutions are parameterized again by the
node number $k$ of the gauge function $w$. 
For any $(r_h,k)$ the behaviour of the functions $w,\sigma$ is 
qualitatively similar to that for regular solutions.
The gauge function $w$ starts from some value $0<w_h<1$ at the horizon and
after $k$ oscillations around zero tends asymptotically to $(-1)^k$.
In this case again one can show that $|w|<1$ everywhere outside the horizon.
In the limit $r_h \to 0$ the event horizon shrinks to zero and the
black hole solutions converges pointwise to the corresponding regular
configuration. In Figure 4 we exhibit the one, two and three nodes solutions of
the $p=2$ model in eight dimensions; similar solutions have been found
for $p=2,d=6$ and $p=3,~d=9,12$.

The stability of solutions in the usual $p=1,\ d=4$ case has been studied
extensively, both perturbatively and at the non-linear level (see the 
discussion in \cite{Volkov:2001tb} and the references therein).
It turns out that all known regular and black hole solutions in that case are
unstable with respect to small spherically symmetric perturbations. The most
obvious indication of the instability comes from the absence of a topological
charge in the YM sector. This is obvious in the $d=4$ case, but when several
members of the YM hierarchy are present like in \cite{BCT,BCHT,BMT}, in some of
these theories a Pontryagin charge is defined, leading to stable solutions.
The stability question in such models has been studied in \cite{BT}. In
the present work however we have restricted to a single (the $p$-th) member of
the YM hierarchy, so that no topological charge can be accommodated in the
case of static solutions.  Indeed, the stability
analysis in \cite{BT} leading to the conclusion that the solutions in
dimensions in which a Pontryagin charge is not defined are unstable
(sphalerons) applies to the models studied in the present Section, and is not
repeated here.

Another aspect of the stability of gravitating YM fields is the
interesting case when a negative cosmological
constant is introduced, in the usual EYM model with $p=1,\ d=4$
\cite{Winstanley:1998sn,Bjoraker:2000qd}. In that case, one can see
from the asymptotic analysis that the  value of
the gauge field function $w(r)$ at infinity is not restricted to $\pm 1$, and
in particular $\lim_{r\to\infty}w(r)=0$ is allowed. This is the asymptotic
value for a static monopole, $i.e.$ the particular solution in question is a
finite energy lump with (topological) monopole charge, rendering it
stable. Unfortunately this property does not persist~\footnote{This can be
seen from the large $r$ asymptotic analysis in the case of $p\ge 2$ models
with negative cosmological constant, which leads exclusively to
$\lim_{r\to\infty}w(r)=\pm 1$.} in the $p\ge 2$ models with negative
cosmological constant, implying that these would be exclusively sphalerons. As
a result we have eschewed an analysis of these.

We close this Section with several comments on the thermodynamic properties
of the black hole solutions. The Hawking temperature of these configurations
can easily be found by using the standard Euclidean method.
For the line element (\ref{metric}), if we treat $t$ as complex, 
then its imaginary part is a coordinate for a non-singular 
Euclidean submanifold iff it is periodic with period 
\begin{eqnarray}\label{period}
\beta=\frac{4 \pi }{N'(r_h)\sigma(r_h)}.
\end{eqnarray}
Then continuous Euclidean Green functions must have this period, so by standard
arguments the Hawking temperature is $T=1/\beta$. For any $(p,d)$ this is found
to decrease with the node number $k$.
To compute the entropy of these solutions, one may use 
the standard relationship between the temperature and the entropy,
$S=\int d{\bf M}/T+S_0$, with $S_0$ a mass-independent constant.
Although further study is necessary, we expect this expression to differ from
the standard one quarter event horizon area value, which holds for $p=1$ EH
gravity only \cite{Myers:1988ze,Wiltshire:1988uq}.


\section{Exact gravity-Yang-Mills solutions in $d=2p+1$}

In the particular dimensions $d=2p+1$, the analogues of \re{lagp},
\be
\label{sis1}
{\cal L}_{(2p+1)}=e\,\left(\frac{\kappa_{p}}{2p}\ \ R_{(p,1)}
+\frac{\tau_p}{2(2p)!}\ \mbox{Tr}\,F(2p)^2\right)\;.
\ee
do not support static finite energy solutions, however, their solutions can be
constructed in closed form.
Since to the best of our knowledge no exact (nontrivial) solution is known 
in the literature for the  
coupled gravity-Yang-Mills equations
\footnote{See, however, the  exact solution with planar symmetry of the $d=4$
EYM equations with a negative cosmological constant $\Lambda=-3$ presented in
\cite{Radu:2004gu}.}, we discuss here the
basic properties of these $d=2p+1$ configurations.
The straightforward generalisation in the presence of
a cosmological constant is presented in Appendix B.

After several redefinitions of the
theory's constants, the effective Lagrangean of this theory reads
\begin{eqnarray}
\label{s1}
L= \sigma \bigg[\frac{d}{dr}(N-1)^p-c r^{1-2p}N (w^2-1)^{2p-2}
\left(\frac{dw}{dr}\right)^{2} \bigg],
\end{eqnarray}
where $c\sim \tau/\kappa$ is a free parameter (in this general case, we do not
fix the sign of $c$).
It is also convenient to redefine the gauge potential according to
\begin{eqnarray}
\label{defa}
a(w)=\int (w^2-1)^{p-1} dw=(-1)^{p+1} {}_2F_1(\frac{1}{2},1-p,\frac{3}{2},w^2),
\end{eqnarray} 
($_2F_1(a,b,c,z)$ being the hypergeometric function),
such that the system (\ref{s1}) admits
the first integral
\begin{eqnarray}
\label{fiw}
a'&\equiv&(w^2-1)^{p-1}w'= \frac{\alpha}{r^{1-2p}N\sigma},
\end{eqnarray} 
where $\alpha$ is an arbitrary real constant.

The metric variables $\sigma$ and $N$ satisfy the equations:
\begin{eqnarray}
\frac{dX^p}{dr}&=&r^{2p-1}\bigg[\frac{c \alpha^2}{(X+1)Y}  \bigg],\label{Xrel}
\\
\frac{dY }{dr}&=&-\frac{2c\alpha^2}{p} \frac{r^{2p-1}}{(X+1)^2X^{p-1}},
\end{eqnarray}
where $X=N-1$, $Y=\sigma^2$.
This implies the relation:
\begin{eqnarray}
\label{rel1a}
\frac{dY}{dX}=-\frac{2Y}{(X+1)},
\end{eqnarray}
which gives
 \begin{eqnarray}
\label{rel1}
Y(X+1)^2=C,
\end{eqnarray}
\newpage
\setlength{\unitlength}{1cm}

\begin{picture}(18,7)
\label{fig5}
\centering
\put(2,0.0){\epsfig{file=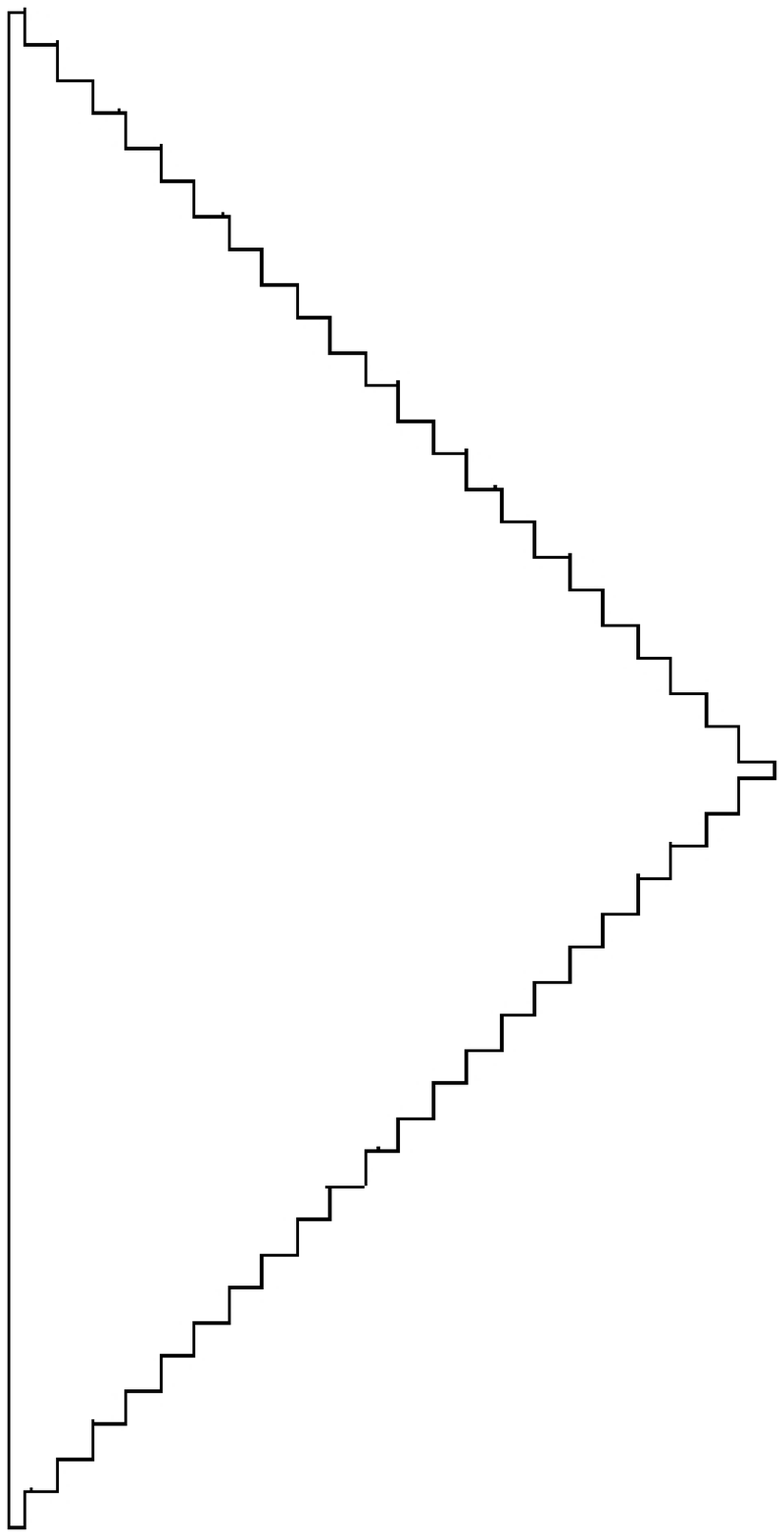,width=10cm}}
\end{picture}
\\
{\small {\bf Figure 5.}
Penrose diagram of the metric (\ref{rel21}) for $c>0$. Wavy lines correspond to
curvature singularities.}
\\
\\
(\textit{i.e.} $N\sigma=C^{1/2}$) and we can set $C=1$, 
without any loss of generality.\footnote{We assume that the integration
constant  $C$ is positive, \textit{i.e.} $\sigma^2>0$.} Replacing this
expression into the $X$-equation one finds the relation:
 \begin{eqnarray}
\label{rel2}
X^p {}_2F_1(p,1,1+p,-X)=\frac{c \alpha^2}{2p}r^{2p}+\beta,
\end{eqnarray}
where $\beta$ is an integration constant.

Expressed in a form which employs $X$ as coordinate, the general solution reads
\begin{eqnarray}
\label{rel3} 
ds^2=f_1(X)dX^2+r^2(X)d\Omega_{2p-1}^2-\frac{dt^2}{X+1},
\end{eqnarray}
where 
\begin{eqnarray}
\label{rel4} 
 f_1(X)&=&
\frac{1}{X+1}\bigg[\frac{pX^{p-1}}{c\alpha^2(X+1)}\bigg]^2r^{2-4p}(X),
 \\
 r(X)&=&\bigg[\frac{2p}{c \alpha^2}\left(X^p {}_2F_1(p,1,1+p,-X)
 -\beta\right)\bigg]^{\frac{1}{2p}}.
\end{eqnarray}
It follows straightforwardly that the expression for the transformed gauge
potential $a$ is 
\begin{eqnarray}
\label{rel5} 
a(X)=a_0+\frac{\alpha}{2p}r^{2p}(X),
\end{eqnarray}
with $a_0$ a constant of integration, which from (\ref{defa}) gives 
the expression of the gauge potential.
One can see that this general solution is not asymptotically flat (as the 
matter fields do not vanish at spacelike infinity). The solution expressed in
the $r$-coordinate takes a simple expression for $p=1$ only
\begin{eqnarray}
\label{rel21}
ds^2&=&{f_0e^{-c\alpha^2 r^2/2}}(dr^2-dt^2)+r^2 d \varphi^2\nonumber\\
w(r)&=&w_0+\frac{\alpha}{2}r^{2},
\end{eqnarray}
where $\alpha$, $f_0$ and $w_0$ are integration constants. In order to 
eliminate the conical singularities in the $(r, \varphi)$ sector we 
require that $f_0=1$. It is straightforward to compute the Kretschmann 
scalar for this metric:
\beqs
K&=&R_{\mu\nu\sigma\rho}R^{\mu\nu\sigma\rho}=
\frac{3c^2\alpha^4e^{c\alpha^2r^2}}{f_0^2}
\eeqs
while the Ricci scalar is proportional to $e^{c\alpha^2r^2/2}$. 
Notice that if $c>0$ these curvature scalars blow up in the limit 
$r\ra\infty$, while apparently the geometry is well behaved and 
becomes regular when $c<0$. The Penrose diagram of the metric (\ref{rel21}) 
in the $c>0$ case is
presented in Figure 5. Each point in the diagram represents a circle 
of radius $r$, as described by the $\varphi$ coordinate. 

While it might be surprising to find null curvature singularities at infinity, 
to understand their presence notice that defining the following null
coordinates:
\beqs
u=t-r, ~~~~~~~ v=t+r,
\eeqs
we can write the metric as:
\beqs
ds^2&=&-f_0e^{-c\alpha^2 (v-u)^2/8}dudv+\frac{1}{4} (v-u)^2 d\varphi^2
\eeqs
while $K$ becomes:
\beqs
K&=&\frac{3c^2\alpha^4e^{c\alpha^2(v-u)^2/4}}{f_0^2}
\eeqs
It is now clear that in the $(r,t)$-plane null rays are represented as
straight lines at $45^o$ with respect to the coordinate lines and furthermore
$K$ diverges as $u\ra\infty$ with $v=const.$ as well as for $v\ra\infty$ with
$u=const.$.

To get a better appreciation of the properties of the $c>0$ $3$-dimensional 
geometry it is of interest to examine in more detail the behaviour of the 
timelike and null geodesics. These will illuminate the nature and the effects 
of the naked curvature singularities at infinity. Since $t$ and $\varphi$ 
are cyclical coordinates we can write down directly the following constants of
motion:
\beqs
e^{-c\alpha^2 r^2/2}\dot{t}=E,~~~~~~~r^2\dot{\varphi}=L
\eeqs
while the radial coordinate $r$ satisfies the equation:
\beqs
\frac{1}{2}\dot{r}^2+V(r)=0
\eeqs
which is the equation of a material point with unit mass and zero energy, 
moving in an effective potential given by:
\beqs
V(r)&=&-\frac{E^2e^{c\alpha^2 r^2}}{2}+\frac{L^2e^{c\alpha^2 r^2/2}}{2r^2}-
\frac{\epsilon e^{c\alpha^2 r^2/2}}{2}
\eeqs
where $\epsilon=-1$ for timelike geodesics, $\epsilon=0$ for null geodesics and
$\epsilon=1$ for spacelike geodesics. Consider first the timelike geodesics, 
\textit{i.e.} $\epsilon=-1$. For non-radial geodesics $L\neq 0$ and for
$E\neq 0$ 
we have $V(r)\ra\infty$ and the region near origin is classically forbidden, 
while in the limit $r\ra\infty$ we have $V(r)\ra-\infty$, which means that 
the particles accelerate so that $|\dot{r}|\ra\infty$ as $r\ra\infty$. For 
radial geodesics $L=0$ and the potential reaches a finite value as $r\ra 0$, 
while again $V(r)\ra-\infty$ as $r\ra\infty$. These properties continue to 
hold for null geodesics. Let us notice that null and even radial timelike 
geodesics can reach spatial infinity in finite intervals of the affine
parameter. Thence the $c>0$ geometry is pathological.

By contrast, if $c<0$, generically $V(r)\ra 0$ when $r\ra\infty$ and therefore 
the region $r\ra\infty$ is indeed at infinity in terms of proper distance and
null right rays cannot reach spatial infinity in finite intervals of the
affine parameter. Then for $c<0$ the geometry is indeed free of any 
curvature singularities. That this is indeed the case can also 
be confirmed by examining the components of the Riemann tensor components 
in an orthonormal frame in the large $r$ limit. 

For either sign of $c$, near origin the spacetime geometry is flat and regular
if $f_0=1$. It is interesting to note that one can also have the case in which 
$f_0<0$, in which case the geometry is time-dependent. However, one can easily
check that the geometry is pathological if $c>0$ and regular for $c<0$.

One should also notice that the $3$-dimensional metric is directly written 
in a Weyl-Papapetrou form. When $c<0$ this will allow us to make an 
interesting connection with a specific time-dependent axially symmetric
geometry of the form:
\beqs
ds^2_4&=&e^{-\sqrt{-2c\alpha^2}t}dz^2+e^{\sqrt{-2c\alpha^2}t}
\bigg[{f_0e^{-c\alpha^2 r^2/2}}(dr^2-dt^2)+r^2 d \varphi^2\bigg],
\eeqs
which is a solution of the vacuum EH field equations in $4$ dimensions,
$R_{ik}=0$.  Upon dimensional reduction along the $z$ coordinate we obtain
precisely the $3$ dimensional geometry (\ref{rel21}) that is now supported by a
time-dependent scalar field $\phi=\sqrt{-2c\alpha^2}t$, which is a solution of
the equations of motion derived from the Lagrangian:
\beqs
{\cal L}_3&=&eR_{(1,1)}-\frac{1}{2}e(\partial\phi)^2
\eeqs
One can now perform the dualisation of the scalar field to obtain a magnetic
2-form field strength and it is easy to check that the final solution
corresponds to the magnetic version of the Reissner-Nordstr{\o}m  solution in
$3$ dimensions \cite{Dias:2002cg}-\cite{Hirschmann:1995he}.

The $4$ dimensional geometry is also free of curvature singularities. If we
perform the  analytical continuations $t\ra iz$, $z\ra it$ and also $c\ra -c$
we obtain the Euclidean version of the $3$ geometry (\ref{rel21}) where now
$c>0$. Its Lorentzian version is obtained if we further analytically continue
$z\ra it$ with the final result that (\ref{rel21}) with $c>0$ is now supported
by a time dependent scalar field whose kinetic term has the wrong sign. This
can be regarded as an indirect confirmation of our previous result that the
geometry (\ref{rel21}) for $c>0$ has pathological properties.

\section{Conclusions}
The general aim of this work was to study the properties of gravitating gauge
field systems whose gravitational part consists of higher order gravitational
curvature terms, e.g. Gauss-Bonnet terms, $i.e.$ necessarily in higher
dimensions. In particular, our scope is limited to static finite energy
solutions, which are also spherically symmetric and asymptotically flat. 
As we know from previous work
that the presence of the usual Einstein-Hilbert (EH, $p=1$ gravity) term masks
the effects of all higher order ($p\ge 2$) terms, we have chosen to study
models featuring only one gravitational term with $p\ge 2$, whose specific
value is chosen according to the dimensionality $d$ of the spacetime. This
still leaves open the choice of the YM terms.

A dual aim in this work was to bring out some general features of gravitating
YM solutions, seen in the original  BK \cite{BK} case with
$d=4$, EH $p=1$ gravity, and usual $p=1$ YM term. This has led us to study
models with $p$-th order gravity and $2p$-th order YM in $d=4p$ dimensions.
This is the content of results in Section 3, where we have verified that all
qualitative properties of the BK solutions are repeated in dimensions
{\it modulo} $4p$ (although nontrivial finite mass
solutions
exists also for  $2p+2\le d \le 4p$). 
Like the BK solution, these are constructed numerically and
are likewise all (unstable) sphalerons. 

As a result of our general considerations we realised that a somewhat different
hierarchy of models, namely those again with $p$-th order gravity and $2p$-th
order YM but now in $d=2p+1$ dimensions, actually supported exact solutions
in closed form. These configurations were discussed in Section 4. Their
properties depend essentially on the sign of the coupling constant
$c\sim \tau/k$. While we expect that the physically meaningful configurations
be described by negative values for $c$, in Section 4 we studied in more
details the $p=1$ case. We found that indeed $c>0$ leads to a pathological
geometry, while for $c<0$ the solution is perfectly regular and we anticipate
these properties to hold for higher values of $p$.
Led on by the work in Section 4, featuring solutions in closed form, we
branched out to introduce a cosmological constant to the models studied in
Section 4. This also led to solutions in closed form, presented in the
Appendices A and B. In Appendix A we considered the gauge decoupled versions
of these models since this turns out to yield a hierarchy of solutions
the first ($p=1$) member of which is the static BTZ
solution~\cite{Banados:1992wn}.
Unfortunately the remarkable geometric features of the $p=1$
solution~\cite{Banados:1992wn} are not repeated in the $p\ge 2$ cases, and a
fairly extensive analysis of this is given in Appendix A. We found that for
generic values of the parameters these solutions are pathological in that they
exhibit naked curvature singularities (that can be hidden inside cosmological
type horizons), however, for special values of the parameters regular
$(a)dS/flat$ backgrounds are obtained. In Appendix B, the closed form solutions
of the models in Section 4 with a non-vanishing $\Lambda$ are presented. The
general solution has two branches and in the $\Lambda\ra 0$ limit only one of
them will survive to give the solution discussed in Section 4. The general form
of these solutions is very complicated and this impedes a general analysis of
their properties. We do expect however that since they have no horizons, they
will generically exhibit naked curvature singularities.

While one might question the physical relevance of the new exact solutions
found in this paper since the form of our gravitational and matter Lagrangians
is non-standard, we take the point of view that given the scarcity of known
non-trivial exact solutions of the YM system coupled to gravity, any new exact
solutions that can be found in closed analytical form might shed some light on 
the properties of such complicated systems. 
Moreover, in literature there have been studied 
modifications of the standard Einstein-Hilbert gravity 
by considering invariant quantities constructed from 
the curvature scalar and/or the Riemann tensor. 
In general it is known that for a gravitational Lagrangian 
constructed out of the metric and the Ricci tensor it is possible 
to perform a conformal transformation (or more generally a Legendre 
transformation) to a metric expressed in the Einstein frame, 
solution of the standard Einstein-Hilbert gravity coupled with 
(exotic) matter fields (see for instance \cite{Koga:1998un,Magnano:1990qu} and
references therein).  The question if a similar reasoning can be applied to the
gravitational lagrangians considered in this paper remains an interesting topic
for further research.
\\
\\
{\bf Acknowledgement}
\\
D.H.T. and E.R. are deeply indepted to Dieter Maison for his help 
and collaboration at the early stage of this work.
C.S. would like to thank Robert Mann for valuable remarks on a draft of this
paper. The work of D.H.T. and E.R. was carried out in the 
framework of Enterprise--Ireland Basic Science Research Project SC/2003/390. 
The work of C.S. was supported by the Natural Sciences and Engineering Council
of Canada.

\vspace{0.8cm}
\appendix
\section{A BTZ hierarchy in $(2p+1)$-dimensions}
\setcounter{equation}{0}
\renewcommand{\theequation}{A.\arabic{equation}}
It is well known that while gravity in $d=3$ spacetime dimensions in the
absence of matter is dynamically trivial, in the presence of a cosmological
constant it supports highly nontrivial asymptotically anti-deSitter (AdS) 
solutions (BTZ)~\cite{Banados:1992wn}.

Our purpose here is to present a hierarchy of gravitational models with
non-vanishing cosmological constant, which support exact solutions in closed
form, proposing a generalisation of the static $d=3$ BTZ solution. 
It turns out that these models are defined in
spacetime dimensions $d=2p+1$ and are described by \re{lagp} with the YM
terms suppressed. This can be written by augmenting \re{lagp},
with YM fields suppressed, by a cosmological constant,
\be
\label{lagexl}
{\cal L}_{\Lambda}={\cal L}_{(2p+1)}\big|_{F=0}-\,(2p+1)!\,e\,\Lambda\,,
\ee
whose $p=1$ member supports the familiar BTZ solution~\cite{Banados:1992wn}.

The exact solutions in question are the static spherically symmetric field
configurations resulting from the imposition of spherical symmetry, given
by the metric Ansatz \re{metric}
in Schwarzschild coordinates, where here $r$ is the $2p$ dimensional
radial coordinate.

Here, we introduce the cosmological constant, $i.e.$
we adopt the system \re{lagexl}. Subjecting the latter to the Ansatz
\re{metric}, after a suitable rescaling we have the reduced Lagrangian
\footnote{Note that  $\Lambda$ has here the opposite sign as compared to the standard choice
in literature} 
\be
\label{2p+1}
L_{(p,1)}^{\Lambda}=\frac{1}{2^{2p-1}}\frac{(d-2)!}{(d-2p-1)!}\,\si\,
\left[\frac{d}{dr}(N-1)^p-r^{2p-1}\,\Lambda\right]\,.
\ee
We immediately find the following solutions
\be
\label{nontrivial}
(N-1)^p=\frac{1}{2p}r^{2p}\,\Lambda+{\rm const.}\qquad,\qquad
p(N-1)^{p-1}\frac{d\si}{dr}=0 \ =>\ \si={\rm const.}
\ee
Notice that for even values of $p$ we have two branches in our solutions. 
We will discuss the global structure of these metrics separately according 
to even or odd values of $p$. 
Also, we will denote the constant that appears in (\ref{nontrivial}) by $M$, 
as this integration constant will be proportional in 
some cases to the mass of a black hole for black hole type spacetimes. 
We will also define $\Lambda=\pm \frac{2p}{\ell^{2p}}$ to simplify notations. 
For $\Lambda>0$ we obtain then in general:
\beqs
N(r)&=&1\pm\left(\frac{r^{2p}}{\ell^{2p}}+M\right)^{\frac{1}{p}},\label{f}
\eeqs
where the minus sign defines a second branch of solutions for even values of
$p$ only. To analyse the case of a negative cosmological constant $\Lambda<0$
one simply analytically continues $\ell\ra i\ell$, while the case $\Lambda=0$
is obtained in the limit $\ell\ra\infty$.

Several particular cases of interest are $p=1$ and $p=2$. 
The former solution corresponds to the celebrated BTZ solution 
\cite{Banados:1992wn} for $\Lambda=\frac{2}{\ell^2}>0$:
\beqs
ds^2&=&-N(r)dt^2+N^{-1}(r)dr^2+r^2d\theta^2\nonumber\\
N(r)&=&\frac{r^{2}}{\ell^2}-M,\label{BTZ}
\eeqs
while $\Lambda<0$ corresponds to $3$-dimensional $dS$ space. If $\Lambda=0$ 
the spacetime is flat and, unless $M=0$, it contains a conical singularity at
origin. For $p=2$ the metric is $5$-dimensional and solves the pure
Gauss-Bonnet equations with a cosmological constant:
\beqs
ds^2&=&-N(r)dt^2+N^{-1}(r)dr^2+r^2d\Omega_{3}^2\nonumber\\
N(r)&=&1\pm\left(\pm\frac{r^{4}}{\ell^4}+M\right)^{\frac{1}{2}}\label{metric5d}
\eeqs
In general, we have two branches of our solutions, which correspond to the
choice of the sign in front of the radical. 
Inside the radical the choice of sign is dictated 
by the sign of the cosmological constant. 
For a negative cosmological constant we have to consider 
strictly positive values for $M$ and, in this case, 
the radial coordinate $r$ will take values only in a finite interval. 
We find, however, that there are curvature singularities at both end points 
of this interval, separated by a black hole type horizon in between. 
For a positive cosmological constant, both positive and negative values of 
the parameter $M$ are allowed as long as the expression under the radical is
positive. We find that there is a naked curvature singularity located at $r=0$,
which is hidden inside a cosmological type horizon. When $M=0$ we obtain 
AdS spacetime as the positive branch solution, 
respectively dS as the negative branch solution. 
If $\Lambda=0$ we obtain a solution containing a naked curvature singularity 
at origin and having a deficit of solid angle.  
As we shall see in the followings, the general solutions in 
higher dimensions exhibit similar properties as the lower-dimensional cases.

The detailed analysis of the global structure of these spacetimes involves a 
discussion of the singularities, horizons and the asymptotic structure. 
We will be mainly interested in curvature singularities and these will 
be identified using the Kretschmann invariant. 
For the spherical symmetric ansatz (\ref{metric}) 
the Kretschman scalar can be written as:
\beqs
K&=&R_{\mu\nu\sigma\rho}R^{\mu\nu\sigma\rho}\nonumber\\
&=&(N'')^2+\frac{2(d-2)}{r^2}(N')^2+\frac{2(d-1)(d-2)}{r^4}(1-N)^2
\eeqs
From the general form of $N(r)$ we can see that there is a curvature
singularity  located at $r=0$ and also at points where
$\frac{r^{2p}}{2p}\Lambda+M=0$. 
Around these curvature singularities the Kretschmann invariant behaves as 
$K\sim O\left(\frac{M^{\frac{2}{p}}}{r^4}\right)$. 
Notice that these metrics are regular only if $M=0$. 

The asymptotic behaviour is controlled by the dominant term in $N(r)$ as
$r\ra\infty$. In general, for odd values of $p$ one can have asymptotically
$(a)dS$ spaces, however, for even values of $p$ we find that we have to
restrict the range of the radial coordinate such that
$\pm\frac{r^{2p}}{\ell^{2p}}+M\geq 0$. 
For $\Lambda=0$ we find that the asymptotic structure is controlled by the 
values of the parameter $M$. If $M=0$ we obtain the flat spacetime. 
If $M\neq0$ then in general we obtain spaces with deficits or surfeits of
solid angle.

To characterise the horizons we use the following definitions \cite{Torii}: 
a horizon located at $r=r_h$ is a null hypersurface with finite curvature, 
such that $N(r_h)=0$. A black hole horizon is defined by the condition
$N'(r_h)>0$; 
a horizon for which $N'(r_h)<0$ and $r_h$ is the largest root of $N(r)$ 
will be called a cosmological horizon. If $N'(r_h)<0$ and $r_h$ is not 
the largest root then $r=r_h$ will define an inner horizon. 
If $N'(r_h)=0$ then $r=r_h$ would correspond to an extreme horizon. 
Notice however that in our solutions this can happen only if $r_h=0$ 
and since $r=0$ is the location of a curvature singularity the spacetime will
be singular.

Consider first spacetime geometries  corresponding to odd values of $p$. 
We will examine the global structure for each value of $\Lambda$ and $M$. 
\begin{itemize}

\item $\Lambda>0$~~~For any value of $M$ the spacetimes are asymptotically
$adS$. If $M>0$, there is a curvature singularity at $r=0$ and since $N(r)$ 
does not vanish for any value of $r$ there are no horizons, hence the 
spacetime contains a globally naked singularity. If $M=0$ then the 
spacetime is $adS$. For $M<0$ we find that $N(r)$ can have a zero at 
$r=r_h$ such that $\frac{r_h^{2p}}{\ell^{2p}}+M=-1$. However, since 
$\frac{r^{2p}}{\ell^{2p}}+M=0$ for a value $r_s>r_h$, then $r=r_s$  
defines a curvature singularity that is not covered by a horizon 
and therefore it corresponds to a globally naked singularity.

\item $\Lambda=0$~~~In this case $N(r)=1+M^{\frac{1}{p}}$ and there 
is a curvature singularity at $r=0$. We exclude from our discussion 
the value $M=-1$ for which $N(r)\equiv 0$. If $M>0$ by rescaling the 
coordinates we can bring the metric in the following form:
\beqs
ds^2&=&-dt^2+dr^2+(1+M^{\frac{1}{p}})r^2d\Omega^2_{d-2}
\eeqs
The spacetime has a surfeit of solid angle as one can see by computing 
the surface area of a sphere with radius $r$ and comparing its value 
with the one calculated in flat spacetime. If $M=0$ we obtain the flat 
spacetime. For negative values of $M$ such that $1+M^{\frac{1}{p}}>0$ 
the spacetime has a deficit of solid angle. If $1+M^{\frac{1}{p}}<0$ then 
$r$ is a timelike coordinate and the spacetime corresponds to a Milne-type 
spacetime having a deficit (or surfeit) of solid angle for $M<-2^p$ 
(respectively $M>-2^p$).

\item $\Lambda<0$~~~If $M<0$ then $N(r)=0$ when $\frac{r^{2p}}{\ell^{2p}}-M=1$ 
and the curvature singularity at $r=0$ is hidden behind a horizon located 
at $r_h=l(1+M)^{\frac{1}{2p}}$. Notice that in order to obtain real values 
for $r_h$ we have to restrict the values of $M$ such that $M>-1$. 
Since $N'(r_h)<0$ this corresponds to a cosmological horizon and 
the spacetime is asymptotically $dS$. For $M=0$ the spacetime becomes 
$dS$ while for $M>0$ the spacetime contains a globally naked singularity 
located at $r_s=lM^{\frac{1}{2p}}$.

\end{itemize}

Let us consider next the geometries corresponding to even values of $p$. 
As we have previously mentioned, there are two branches of solutions 
and we shall discuss each branch separately. The negative branch has:
\beqs
N(r)&=&1-\left(\pm\frac{r^{2p}}{\ell^{2p}}+M\right)^{\frac{1}{p}}
\eeqs
Here the upper sign corresponds to a positive cosmological constant
$\Lambda>0$, the lower sign corresponds to $\Lambda<0$, while the case
$\Lambda=0$ is obtained in the limit $\ell\ra\infty$. Notice that we have to
restrict the values of the parameters $M$, $l$ and of the radial coordinate
$r$ such that $\pm\frac{r^{2p}}{\ell^{2p}}+M\geq 0$.
\begin{itemize}

\item $\Lambda>0$~~~For $M>0$ the spacetime is asymptotically $dS$ and it 
has a curvature singularity at $r=0$, hidden inside a cosmological horizon 
located at $r_h=l(1-M)^{\frac{1}{2p}}$. Notice that in order to obtain real 
values for $r_h$ we must restrict the values for $M$ such that $M<1$. For 
$M=1$ then $r_h=0$ and the spacetime contains a globally naked singularity. 
If $M>1$ there is a globally naked singularity located at $r=0$. If $M=0$ 
we obtain $dS$. For $M<0$ we find a singularity at $r=0$ inside of a 
cosmological horizon located at $r_h=\ell(1-M)^{\frac{1}{2p}}$.

\item $\Lambda=0$~~~In order to avoid complex values for
$N(r)=1-M^{\frac{1}{2p}}$ 
we must require that $M\geq 0$. If $M=0$ we obtain Minkowski spacetime. 
If $M>0$ the spacetime contains a globally naked singularity at $r=0$ and 
it has a deficit of solid angle.

\item $\Lambda<0$~~~If $M\leq0$ we find that $f(r)$ takes complex values. 
If $M>0$ we must also restrict the values of the radial coordinate to 
a finite interval such that $M-\frac{r^{2p}}{\ell^{2p}}\geq 0$. 
There exists a black hole type horizon at $r_h=\ell(M-1)^{\frac{1}{2p}}$ only
if $M>1$. However, at both endpoints of the radial coordinate there are
curvature singularities.  For $M=1$ there is a naked curvature singularity at
$r=0$ and also another one at $r=\ell$.

\end{itemize}

Consider next the positive branch solutions. In this case we have:
\beqs
N(r)&=&1+\left(\pm\frac{r^{2p}}{\ell^{2p}}+M\right)^{\frac{1}{p}}
\eeqs
The analysis is very similar with the one performed for odd values of $p$. 
There exists however an extra restriction that
$\pm\frac{r^{2p}}{l^{2p}}+M\geq 0$. 
The equality sign corresponds to a curvature singularity location. 
We find then that these spacetimes are generically singular unless 
$M=0$ and in the cases where there are regular horizons, 
these correspond to cosmological horizons such that the 
spacetimes still contain globally naked singularities.

\section{ $D=2p+1$ gravity-Yang-Mills solutions with $\Lambda\neq 0$}
\setcounter{equation}{0}
\renewcommand{\theequation}{B.\arabic{equation}}
For $p=1,~d=4$, the EYM equations with a cosmological term present solutions
with very different properties as compared to the $\Lambda=0$ case 
\cite{Winstanley:1998sn,Bjoraker:2000qd}.
Their $d>4$ generalisations in Einstein gravity with higher terms in the
Yang--Mills hierarchy have been discussed recently in \cite{Radu:2005mj}.
It is therefore natural to consider solutions of the 
$p$-th gravity-Yang-Mills system (\ref{sis1}) 
in the presence of a cosmological constant, 
in $d=2p+1$ dimensions.

Using the same notations as in Section 4, 
the reduced Lagrangean of this system reads
\begin{eqnarray}
\label{s11}
L= \sigma \bigg[\frac{d}{dr}(N-1)^p-c r^{1-2p}N (w^2-1)^{2p-2}
\left(\frac{dw}{dr}\right)^{2}-\Lambda r^{2p-1}\bigg],
\end{eqnarray}
the resulting field equations being solved by using the same techniques as
in the $\Lambda=0$ case.
 
Employing the same variables, $X=N-1$ and $Y=\sigma^2$, we find the same
$\sigma$-equation, while the function $N$ satisfy the equation 
\begin{eqnarray}
\frac{dX^p}{dr}&=&r^{2p-1}\bigg[\frac{c \alpha^2}{(X+1)Y}+\Lambda\bigg],
\label{Xrel1}
\end{eqnarray} 
the gauge field equations still admitting the first
integral (\ref{fiw}).

Therefore (\ref{rel1a}) is generalized to
\begin{eqnarray}
\label{rel1b}
\frac{dY}{dX}=-\frac{2Y}{(X+1)\big(1+\lambda Y(X+1)\big)},
\end{eqnarray}
where $\lambda=\Lambda/c\alpha^2$. 
If we define a new function $Z=(X+1)Y$, we find that (\ref{rel1b}) can be
written as:
\beqs
\frac{dZ}{dX}&=&\frac{Z}{X+1}\frac{\lambda Z-1}{\lambda Z+1}
\eeqs
and upon integration we obtain:
\beqs
\frac{(\lambda Z-1)^2}{Z}&=&\frac{X+1}{C}
\label{Z}
\eeqs
where $C$ is an integration constant. 
This equation  has now two solutions, denoted by $Z_{\pm}$ and given by:
\beqs
\lambda Z_{\pm}&=&\frac{\sqrt{X+c_1}\mp\sqrt{X+1}}
{\sqrt{X+c_1}\pm\sqrt{X+1}},
\eeqs
where $c_1=4\lambda C+1$. In consequence, we obtain two branches for our
solution:
\begin{eqnarray}
\label{rel6} 
Y_{\pm}=\frac{c\alpha^2}{ \Lambda (1+X)} \frac{\sqrt{X+c_1}\pm\sqrt{X+1}}
{\sqrt{X+c_1}\mp\sqrt{X+1}},
\end{eqnarray}
  Notice that when taking the limit $\Lambda\ra 0$ only the 
  negative branch will survive and we obtain precisely 
  the solution given in (\ref{rel1}). The positive branch 
  solution has no asymptotically flat limit. Replacing 
  these expressions in (\ref{Xrel}), we arrive at a relation similar to
 (\ref{rel2}):
\begin{eqnarray}
\label{rel7}
\frac{1}{2p}r^{2p}+\beta=\frac{p}{2\Lambda}F_{\pm}(X),
\end{eqnarray}
where 
\begin{eqnarray}
\label{rel8}
 F_{\pm}(X)=\int X^{p-1}\left(1\pm\sqrt{\frac{X+1}{X+c_1}}\right)dX.
\end{eqnarray}
It appears that it is not possible to find a general expression of 
this integral. Several particular cases of potential interest 
in which we can integrate (\ref{rel8}) are:
\begin{eqnarray}
\label{rel9}
\nonumber
 F_{\pm}(X)&=&X\pm\sqrt{(X+1)(X+c_1)}\mp(c_1-1)\log (\sqrt{X+1}+\sqrt{X+c_1}),
~~~{\rm for}~~~p=1.
 \\
F_{\pm}(X)&=&\frac{1}{8}
(4X^2\pm2\sqrt{(X+1)(X+c_1)}(1-3c_1+2X)
\\
\nonumber
&&\pm(c_1-1)(3c_1+1)
\log (1+c_1+2X+2\sqrt{(X+1)(X+c_1)}))~~~{\rm for}~~~p=2.
\end{eqnarray}

The general solution takes a simpler form when expressed using $X$ as
coordinate with
\begin{eqnarray}
\label{rel12}
 r(X)=\left(\frac{p^2}{\Lambda}F_{\pm}(X)-2p\beta\right)^{\frac{1}{2p}}.
\end{eqnarray}
Therefore, the general metric of the $\Lambda\neq 0$ solution is given by
\begin{eqnarray}
\label{rel13}
ds^2=g_1(X)dX^2+r^2(X)d\Omega_{d-2}^2-g_2(X)dt^2
\end{eqnarray}
with
\begin{eqnarray}
\label{rel14}
g_1(X)&=&\Bigg[\frac{pr^{-2p-1}(X)}{2\Lambda }X^{p-1}
\left(1\pm\sqrt{\frac{X+1}{X+c_1}}\right)\Bigg]^2\frac{1}{X+1},
\\
\nonumber
g_2(X)&=&\frac{c\alpha^2}{ \Lambda} \frac{\sqrt{X+c_1}\pm\sqrt{X+1}}
{\sqrt{X+c_1}\mp\sqrt{X+1}}.
\end{eqnarray}
From the above form of the general solution, notice that we have to restrict
the $X$ coordinate such that $X\geq -1$ and that there are no horizons.
However $g_1(X)$ will blow up as $X\ra -1$ and therefore we conclude that in
general such spaces will have pathological features.

\newpage


\end{document}